\documentclass[twocolumn,aps,pre,amsmath,amssymb]{revtex4-1}
\usepackage{graphicx}
\usepackage{dcolumn}
\usepackage{bm}

\usepackage{multirow}

\usepackage{amssymb}
\usepackage{amsmath}
\usepackage{graphicx}
\usepackage{xcolor}
\usepackage[colorlinks,citecolor=blue,linkcolor=red,urlcolor=blue]{hyperref}
\usepackage{CJK}
\usepackage{indentfirst}
\usepackage{amsmath}
\usepackage{cases}

\begin{document}
\title{Imaginary-time relaxation quantum critical dynamics in two-dimensional dimerized Heisenberg model}
\author{Jia-Qi Cai$^{1}$}
\author{Yu-Rong Shu$^{2}$}
\author{Xue-Qing Rao$^{1}$}
\author{Shuai Yin$^{1}$ }
\email{yinsh6@mail.sysu.edu.cn}
\affiliation{$^1$Guangdong Provincial Key Laboratory of Magnetoelectric Physics and Devices, School of Physics, Sun Yat-sen University, Guangzhou 510275, China}
\affiliation{$^2$School of Physics, Guangzhou University, Guangzhou 510275, China}

\date{\today}
\begin{abstract}
We study the imaginary-time relaxation critical dynamics of the N\'{e}el-paramagnetic quantum phase transition in the two-dimensional (2D) dimerized $S=1/2$ Heisenberg model. We focus on the scaling correction in the short-time region. A unified scaling form including both short-time and finite-size corrections is proposed. According to this full scaling form, improved short-imaginary-time scaling relations are obtained. We numerically verify the scaling form and the improved short-time scaling relations for different initial states using projector quantum Monte Carlo algorithm.
\end{abstract}
\maketitle

\section{Introduction}
Quantum phase transitions (QPTs) describe nonanalytic changes between different ground states of many-body systems~\cite{Sachdevbook}. Although QPTs are governed by quantum fluctuations at zero temperature, they can remarkably affect the finite-temperature phase diagram, giving rise to a variety of exotic behaviors in the famous quantum critical regime as exhibited in a wide range of strongly correlated systems~\cite{Sachdevbook,Sondhi1997rmp}. Thus the QPTs have received considerable attentions from both theoretical and experimental aspects. Among various models of QPTs, the $S=1/2$ Heisenberg antiferromagnet has attracted enormous investigations~\cite{Chakravarty1988prl,Huse1988prl,Singh1989prb,Millis1993prl,Chubukov1994prb,Troyer1996prb,Troyer1998prl,Matsumoto2001prb,Sandvik2006prb,Giamarchi2008natphys,Sachdev2008natphys,Sandvik2010review,Merchant2014natphy,Mila2015prb,Wenzel2008prl,Meng2015prb,Nvsen2018prl,Siqimiao2018prb,Jiang2018prb,Jiang2020prb,Sandvik1994prl}, not only because it is one of the typical quantum models whose ordered phase can spontaneously break continuous symmetry, but also owing to its close relation to strongly-correlated materials such as the cuprate superconductors~\cite{McMorrow2001prl,Sachdev2008natphys,RManousakis1991amp,Vojta2007rmp}.

Recent investigations on QPTs are increasingly focusing on their nonequilibrium dynamics, because the interplay between the divergent correlation time scale and the breaking of the translation symmetry in time direction can trigger lots of intriguing universal dynamic behaviors, which usually go beyond the conventional scheme of equilibrium QPTs~\cite{Polkovnikov2011rmp,Dziarmaga2010review,Masahito2013rmp,Rigol2016review,Mitra2018arcmp,Werner2013prl,Sieberer2013prl,heyl2013prl,Yin2019prl,Yin2021prb,Berges2021rmp,Berges2004prl,Mitra2015prb,Halimeh2023prb,Marino2022review,JianLi2023arxiv}.  For example, in equilibrium, quantum criticality in $d$-dimension can usually be mapped into the corresponding classical criticality in $(d+1)$-dimension via the path-integral formulation~\cite{Sachdevbook,Sondhi1997rmp}. In contrast, for the nonequilibrium case, there is no similar mapping between quantum and classical critical dynamics. Besides the theoretical novelty, intriguing nonequilibrium critical phenomena have been found in various experiments~\cite{Clark2016science,Du2023,Navon2015science,Lamporesi2013,Nicklas2015prl,Jurcevic2017prl}. Moreover, quantum critical dynamics also has important applications in preparing and characterizing various exotic quantum phases in fast-developing quantum devices~\cite{Weinberg2020prl,moore2022prb,king2023nature,Semeghini2021science,Ebadi2021}.

Aside from the real-time dynamics, imaginary-time dynamics in quantum systems is also of great interest and significance. Because of its dissipative nature, the imaginary-time evolution usually works as a routine unbiased method to determine the ground state, not only widely used in numerical simulations, such as the time-evolving block decimation~\cite{Vidal2004prl,Vidal200prl}, tensor network~\cite{Vidal2008prl,jiang200prl}, and quantum Monte Carlo (QMC)~\cite{Sandvik2010review,AssaadReview,Li2019Review}, but also in rapidly developing quantum computers~\cite{Motta2020naturephyscis,Nishi2021njp,Pollmann2021prxq}. Near a quantum critical point, it was shown that universal scaling behaviors appear not only in the long-time equilibrium stage, but also in short-time relaxation stage after a transient time scale~\cite{Yins2014prb,Yins2014pre}. So far, the short-imaginary-time quantum critical dynamics has been studied in various quantum systems, including the quantum Ising model~\cite{Yins2014prb,Yins2014pre,Shu2017prb,Shu2020prb}, deconfined quantum criticality~\cite{Yin2022prl,Yin2022prb}, and strongly-correlated Dirac systems~\cite{Yin2023arxiv1}, providing an abundance of intriguing perspectives in the field of quantum criticality. In addition, the short-imaginary-time scaling behavior has been detected in an experimental platform of a noisy intermediate-scale quantum computer~\cite{Zhang2023}. Moreover, the short-imaginary-time critical dynamics also shows its power in determining the critical properties with high efficiency, circumventing difficulties induced by critical slowing down and divergent entanglement entropy encountered in conventional methods based on equilibrium scaling~\cite{Yins2014prb,Yins2014pre,Shu2017prb,Shu2020prb,Yin2022prl,Yin2022prb,Yin2023arxiv1,Zhang2023}. However, up to now, the short-imaginary-time scaling property has not been explored in the quantum Heisenberg universality class yet.

\begin{figure}[tbp]
\centering
  \includegraphics[width=\linewidth,clip]{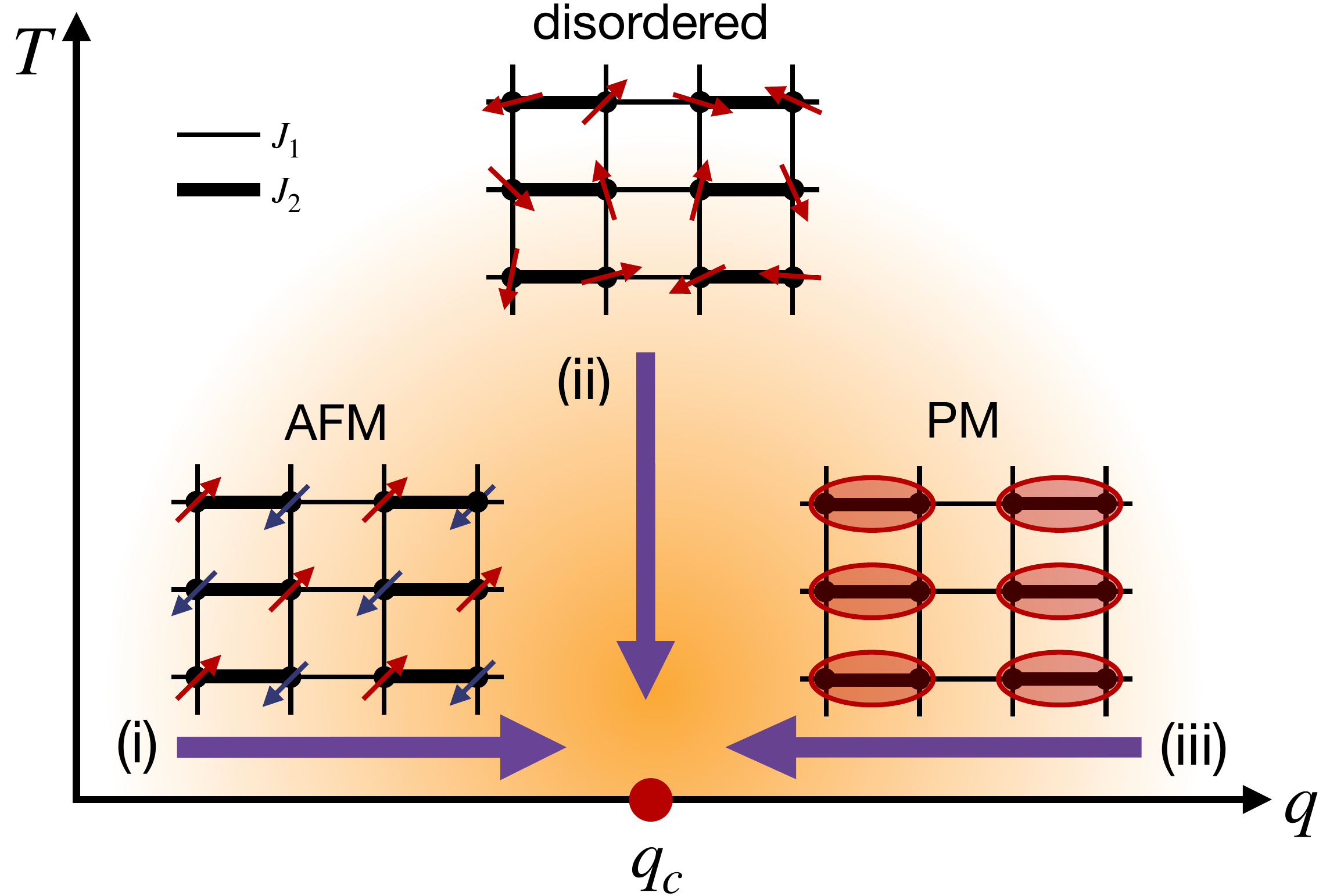}
  \vskip-3mm
  \caption{Sketch of the phase diagram and the quench protocol in imaginary-time with different initial states. $J_1$ and $J_2$ are the antiferromagnetic coupling on bonds $\langle ij \rangle$ (thin) and $\langle ij \rangle'$ (thick), respectively. The initial states are prepared as (i) the N\'{e}el antiferromagnetic phase, (ii) the disordered state, and (iii) the dimerized paramagnetic state. All states correspond to the fixed points of the initial states under the renormalization group transformation.
  }
  \label{figure1}
\end{figure}

In this paper, we explore the imaginary-time relaxation critical dynamics of the Heisenberg universality class in the QPT between N\'{e}el antiferromagnetic (AFM) and quantum paramagnetic (PM) states in the two-dimensional dimerized Heisenberg model with inter- and intra-dimer couplings $J_1$ and $J_2$, as illustrated in Fig.~\ref{figure1}. We find that the relaxation dynamics of this model exhibits scaling behaviors with remarkable short-time scaling corrections. A unified scaling form including both short-time and finite-size corrections is developed. From this scaling function, short-imaginary-time scaling properties with scaling correction included can be inferred. For different initial states, we find that the relaxation dynamics in the imaginary-time direction can be well described by this scaling form. The short-time scaling relations are also verified numerically. Our present work not only reveals the imaginary-time relaxation critical dynamics in the Heisenberg universality class, but also provides a systematic scaling analysis on the scaling corrections in the time direction, which be generalized to other kinds of nonequilibrium critical dynamics.


The rest of the paper is arranged as follows. The dimerized Heisenberg model is introduced in Sec.~\ref{secmodel}. Then, in Sec.~\ref{sectheory}, after a brief review on the original short-imaginary-time scaling theory in Sec.~\ref{sectheory1}, the scaling theory with short-time corrections is developed in Sec.~\ref{sectheory2}. The main numerical results are shown in Sec.~\ref{Results}. At last, a summary is given in Sec.~\ref{summary}.

\section{\label{secmodel}Model}
The Hamiltonian of the $2$D dimerized Heisenberg model reads~\cite{Chakravarty1988prl,Huse1988prl,Singh1989prb,Millis1993prl,Chubukov1994prb,Sachdev2008natphys}
\begin{equation}
H=J_1 \sum_{\langle ij \rangle}\mathbf{S}_i\cdot\mathbf{S}_j+J_2 \sum_{\langle ij \rangle'}\mathbf{S}_i\cdot\mathbf{S}_j,
\label{eq:hamiltonian}
\end{equation}
in which $\mathbf{S}_i=(1/2)(\sigma_x,\sigma_y,\sigma_z)$ denotes the spin-$1/2$ operator at site $i$, $J_1$ and $J_2$ are the antiferromagnetic coupling constants defined on the bonds $\langle ij \rangle$ and $\langle ij \rangle'$, respectively, as illustrated in Fig.~\ref{figure1}. When $q\equiv J_2/J_1\approx 1$, the ground state of Eq.~(\ref{eq:hamiltonian}) hosts the N\'{e}el AFM order~\cite{Sandvik2010review,Nvsen2018prl}, characterized by the order parameter ${\bf M}\equiv (1/L^2)\sum_r (-1)^{r_x+r_y}\mathbf{S}_r$. In contrast, when $q>q_c=1.90951(5)$~\cite{Nvsen2018prl}, the ground state changes to the paramagnetic (PM) state. It was shown that the dimerized Heisenberg model can be mapped to a nonlinear sigma model with an irrelevant Berry phase term and its criticality is well described by the Heisenberg $O(3)$ universality class~\cite{Chakravarty1988prl,Chubukov1994prb,Sachdev2008natphys}. This claim has been verified with scrutiny by numerical simulations via efficient quantum Monte Carlo methods~\cite{Sandvik2010review}.

\section{\label{sectheory}Scaling theory in short-imaginary-time quantum critical dynamics}
In this section, after briefly reviewing the short-imaginary-time scaling theory in Sec.~\ref{sectheory1}, we generalize this theory to include the short-time and finite-size scaling corrections, as illuminated in Sec.~\ref{sectheory2}.

\subsection{\label{sectheory1}Brief review on short-imaginary-time quantum critical dynamics}
The imaginary-time evolution of a quantum state $|\psi(\tau)\rangle$ is described by the imaginary-time Schr\"{o}dinger equation
\begin{equation}
 -\frac{\partial}{\partial \tau} |\psi(\tau)\rangle=H|\psi(\tau)\rangle,
\label{eq:dynamiceq}
\end{equation}
imposed additionally by the normalization condition $\langle\psi(\tau)|\psi(\tau)\rangle=1$. Its formal solution is $|\psi(\tau)\rangle=Z \exp(-\tau H)|\psi(0)\rangle$, in which $|\psi(0)\rangle$ is the initial wave vector and $Z\equiv 1/\|\exp(-\tau H)|\psi(0)\rangle\|$ is the normalization factor. The expectation value of an operator $\mathcal{\hat{Q}}$ at $\tau$ is then given by
\begin{equation}
 \mathcal{Q}(\tau)=\langle\psi(\tau)|\mathcal{\hat{Q}}|\psi(\tau)\rangle
\label{eq:qtau}
\end{equation}

For a gapped quantum system with an arbitrary initial state $|\psi(0)\rangle$, in which $|\psi(0)\rangle$ is assumed to have some overlap with the ground state, the wave function $|\psi(\tau)\rangle$, evolving according to Eq.~(\ref{eq:dynamiceq}), will fast decay to the ground state after a time scale $\tau_\Delta\sim 1/\Delta$ with $\Delta$ being the gap. Based on this, the imaginary-time evolution according to Eq.~(\ref{eq:dynamiceq}) provides an effective method to find the ground state numerically~\cite{Vidal2004prl,Vidal200prl,Vidal2008prl,jiang200prl,Sandvik2010review,AssaadReview,Li2019Review}. In contrast, when the system is at its critical point, $\Delta\rightarrow 0$ in the thermodynamics limit and thus $\tau_\Delta$ diverges. This reflects the critical slowing down in quantum phase transitions.

Associated with the divergence of $\tau_\Delta$, when the quantum system is near its critical point, nonequilibrium scaling behaviors appear in the imaginary-time relaxation process. It was shown that the scaling form of quantum imaginary-time dynamics is similar to that of classical short-time critical dynamics since both of them feature the dissipative nature~\cite{Yins2014prb,Yins2014pre,Janssen1989,Albano2011iop,Lizb1995prl,Zheng1996prl}. In the imaginary-time relaxation process, the general scaling form for an operator $\mathcal{Q}$ satisfies~\cite{Yins2014prb,Yins2014pre}
\begin{equation}
\label{eq:general}
\mathcal{Q}(\tau,\delta,L,\{Y\})=\tau^{\kappa/z}f_Q(\delta \tau^{1/\nu z},\tau L^{-z}, \{Y_0 \tau^{-y_0/z}\}),
\end{equation}
in which $\kappa$ is the critical exponent related to $\mathcal{Q}$, $\delta=q-q_c$ is the distance to the critical point and has a dimension of $1/\nu$ with $\nu$ being the correlation length exponent, $z$ is the dynamic exponent and ${Y_0}$ with its exponent $y_0$ denotes the relevant initial information. In the long-time limit, i.e., $\tau\gg L^{-z}$ and $\tau \gg |\delta|{-\nu}$, $Y_0$ vanishes and the usual equilibrium finite-size scaling form is recovered.

In particular, when the initial state corresponds to the fixed point of renormalization group transformation, such as the completely ordered and disordered initial states, the term of $Y_0$ can hide away in Eq.~(\ref{eq:general}). In this way, the scaling form~(\ref{eq:general}) becomes~\cite{Yins2014prb,Yins2014pre,Janssen1989,Albano2011iop,Lizb1995prl,Zheng1996prl}
\begin{equation}
\label{eq:general1}
\mathcal{Q}(\tau,\delta,L)=\tau^{\kappa/z}f_{Q1}(\delta \tau^{1/\nu z},\tau L^{-z}).
\end{equation}
Note that in this case the detailed function of $f_{Q1}$ still implicitly depends on $Y_0$. For example, from the completely ordered initial state, the evolution of the square of the order parameter, $M^2$, at $\delta=0$ satisfies~\cite{Yins2014prb}
\begin{equation}
\label{eq:op1}
M^2(\tau,L)=\tau^{-2\beta/\nu z}f_{M1}(\tau L^{-z}),
\end{equation}
in which $\beta$ is the order parameter exponent defined as $M\propto |\delta|^{\beta}$ in the ordered phase. In the short-time stage, $f_{M1}(\tau L^{-z})$ tends to a constant and
\begin{equation} 
 \label{eq:op1b}
 M^2\propto \tau^{-2\beta/\nu z}, 
\end{equation}
whereas in the long-time stage, $f_{M1}(\tau L^{-z})\propto (\tau L^{-z})^{2\beta/\nu z}$ and scaling form restores to $M^2\propto L^{-2\beta/\nu}$~\cite{Sandvik2010review}, which is the leading term of the finite-size scaling form
\begin{equation}
\label{eq:op1a}
M^2(\tau,L)=L^{-2\beta/\nu}f_{M2}(\tau L^{-z}).
\end{equation}

In addition, from the completely disordered initial state, $M^2$ obeys~\cite{Albano2011iop,Yin2022prl}
\begin{equation}
\label{eq:op2}
M^2(\tau,L)=L^{-d}\tau^{-2\beta/\nu z+d/z}f_{M3}(\tau L^{-z}),
\end{equation}
for $\delta=0$. In Eq.~(\ref{eq:op2}), $M^2\propto L^{-d}$ in the leading term comes from the the central limit theorem when the correlation length is smaller than $L$~\cite{Albano2011iop}. In the short-time stage, $f_{M3}(\tau L^{-z})$ tends to a constant and 
\begin{equation}
\label{eq:op2a}
M^2\propto L^{-d}\tau^{-2\beta/\nu z+d/z},
\end{equation} 
whereas in the long-time stage, $f_{M3}(\tau L^{-z})\propto (\tau L^{-z})^{2\beta/\nu z-d/z}$, giving rise to $M^2\propto L^{-2\beta/\nu}$ and Eq.~(\ref{eq:op1a}).

\subsection{\label{sectheory2}Scaling corrections in short-imaginary-time quantum critical dynamics}
The above scaling analyses only consider the leading contributions of the relevant scaling variables. However, scaling corrections from the subleading contributions are also very important in describing quantum criticality. For example, it was shown that the finite-size scaling correction plays significant roles in determining critical properties for practical numerical simulations~\cite{Sandvik2010review}. Moreover, figuring out scaling corrections also provides strong evidences to clarify the universality classes of QPTs~\cite{Sandvik2010review,Nvsen2018prl,Meng2015prb,Wenzel2008prl,Wessel2023prb}. Previous investigations mainly focus on the scaling corrections from finite-size effects~\cite{Sandvik2010review}. For the nonequilibrium critical dynamics, the time is an intrinsic variable, such that the scaling correction from the time direction is quite essential and should be taken care carefully.

For simplicity, in the following, we shall consider the cases for which $\delta=0$ and initial states are at their fixed points under scale transformation. We start with the general scaling form of a quantity $\mathcal{Q}$
\begin{equation}
\label{eq:general2}
\mathcal{Q}(\tau,L)=L^{\kappa}f_{Q3}(\tau L^{-z},L^{-\omega_L},\tau^{-\omega_\tau}),
\end{equation}
in which $L^{-\omega_L}$ is the usual finite-size scaling correction~\cite{Sandvik2010review,Jiang2018prb} and $\omega_L$ is the correction exponent, and $\tau^{-\omega_\tau}$ represents the short-time scaling corrections with $\omega_\tau$ the correction exponent. In Eq.~(\ref{eq:general2}), we assume that $\omega_\tau=\omega_L$ and both of them are denoted as $\omega$. This assumption is based on the fact that the critical theory of the quantum Heisenberg model has the Lorentz symmetry~\cite{Chakravarty1988prl,Huse1988prl,Singh1989prb,Millis1993prl,Chubukov1994prb,Sachdev2008natphys} and will be verified by the numerical results in the next section.

However, directly using the full ansatz Eq.~(\ref{eq:general2}) is certainly unpractical since the detailed form of scaling function $f_{Q3}$ is unknown. Here, we propose that Eq.~(\ref{eq:general2}) can be approximated as
\begin{equation}
\label{eq:general3}
\mathcal{Q}(\tau,L)=L^{\kappa}(1+b_QL^{-\omega})f_{Q4}[\tau L^{-z}(1+a_Q\tau^{-\omega})],
\end{equation}
in which $b$ is the coefficient of finite-size correction and equals its equilibrium value, and $a$ is the coefficient of the short-time correction. Both $a$ and $b$ depend on the $\mathcal{Q}$. In addition, $a$ should also depend on the initial state.

The properties of Eq.~(\ref{eq:general3}) are discussed as follows. First, in the long-time limit, i.e., $\tau\rightarrow \infty$, $\tau^{-\omega}$ vanishes and $f_{Q4}(\tau L^{-z})$ tends to a constant. Accordingly, Eq.~(\ref{eq:general3}) is reduced to
\begin{equation}
\label{eq:general4}
\mathcal{Q}(\tau,L)\propto L^{\kappa}(1+b_QL^{-\omega}),
\end{equation}
which is consistent with the usual finite-size scaling relation with finite-size scaling correction included~\cite{Sandvik2010review,Jiang2018prb,Nvsen2018prl}. 

\begin{figure*}[tbp]
\centering
  \includegraphics[width=\linewidth,clip]{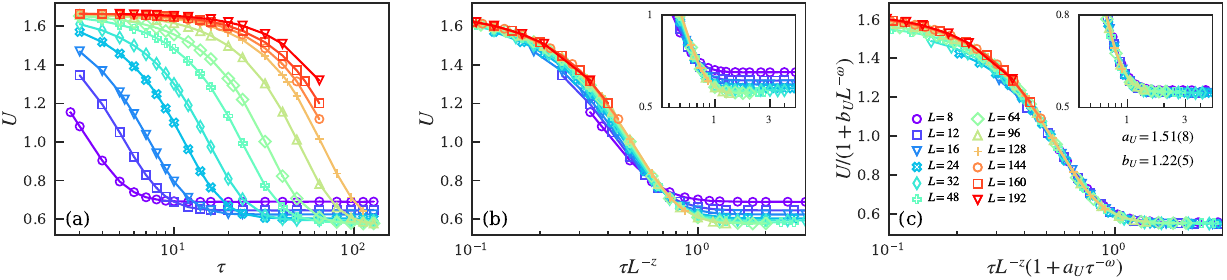}
  \vskip-3mm
  \caption{(a) Evolution of the Binder ratio from the completely ordered initial state for different lattice sizes at the critical point is shown. (b) When $\tau$ is rescaled as $\tau L^{-z}$, deviations still appear for the rescaled curves of $U$ versus $\tau$. (c) Rescaled curves collapse well when both short-time and finite-size scaling corrections are added. Note that in (c), the correction exponents for both time and size are set as $\omega=0.78$. Logarithmic scale is used in x-axis.
  }
  \label{figure2}
\end{figure*}

Second, to reveal the short-time scaling relations, we set $\tau L^{-z}(1+a_Q\tau^{-\omega})=c$, namely, $L=[\tau (1+a_Q\tau^{-\omega})/c]^{1/z}$. By substituting this equation into the leading term of Eq.~(\ref{eq:general3}), we obtain
\begin{eqnarray}
\mathcal{Q}(\tau,L)&=&\tau^{\kappa/z}(1+a_Q\tau^{-\omega})^{\kappa/z}\nonumber    \\
 &\times&(1+b_QL^{-\omega})f_{Q5}[\tau L^{-z}(1+a_Q\tau^{-\omega})], \label{eq:general5}
\end{eqnarray}
in which $c^{1/z}$ has been absorbed into $f_{Q5}$. From Eq.~(\ref{eq:general5}), one finds that for large $L$ in the short-time stage, the evolution of $\mathcal{Q}$ satisfies
\begin{equation}
\label{eq:general6}
\mathcal{Q}(\tau)\propto \tau^{\kappa/z}(1+a_Q\tau^{-\omega})^{\kappa/z}.
\end{equation}

We will illustrate the short-imaginary-time scaling theory with scaling corrections for different quantities. For example, at the critical point $\delta=0$, with the initial state corresponding to its fixed point, the evolution of the dimensionless Binder cumulant, defined as $U\equiv \frac{5}{2}(1-\frac{5\langle M\rangle^4}{3\langle M^2\rangle^2}$), should satisfy
\begin{equation}
\label{eq:binder}
U(\tau,L)=(1+b_UL^{-\omega})f_{U}[\tau L^{-z}(1+a_U\tau^{-\omega})],
\end{equation}
according to Eq.~(\ref{eq:general3}).

In addition, the square of the order parameter $M^2$ should obey
\begin{equation}
\label{eq:op3}
M^2(\tau,L)=L^{-2\beta/\nu}(1+b_ML^{-\omega})f_{M4}[\tau L^{-z}(1+a_M\tau^{-\omega})].
\end{equation}
In particular, for an ordered initial state, the evolution of $M^2$ should follow the scaling form
\begin{eqnarray}
M^2(\tau,L)&=&\tau^{-2\beta/\nu z}(1+a_M\tau^{-\omega})^{-2\beta/\nu z}\nonumber    \\
 &\times&(1+b_ML^{-\omega})f_{M5}[\tau L^{-z}(1+a_M\tau^{-\omega})], \label{eq:op4}
\end{eqnarray}
according to Eq.~(\ref{eq:general5}). Note that here the dependence of $a_M$ on the initial states is not explicitly shown to avoid complex labels. For large $L$, Eq.~(\ref{eq:op4}) indicates that in the short-time stage
\begin{eqnarray}
M^2(\tau)\propto\tau^{-2\beta/\nu z}(1+a_M\tau^{-\omega})^{-2\beta/\nu z}, \label{eq:op5}
\end{eqnarray}
according to Eq.~(\ref{eq:general6}). Comparing with Eq.~(\ref{eq:op1b}), one finds that a correction factor has been multiplied. In the next section, we will find that the factor is crucial in describing the short-imaginary-time relaxation dynamics of model~(\ref{eq:hamiltonian}).


Moreover, for a disordered initial state, $M^2$ should obey
\begin{eqnarray}
M^2(\tau,L)&=&L^{-d}\tau^{-2\beta/\nu z+d/z}(1+a_M\tau^{-\omega})^{-2\beta/\nu z+d/z}\nonumber    \\
 &\times&(1+b_ML^{-\omega})f_{M6}[\tau L^{-z}(1+a_M\tau^{-\omega})]. \label{eq:op8}
\end{eqnarray}
Note that the leading term of $L^{-d}$ is assumed to be not affected by the scaling correction, since $M^2(\tau, L)\propto L^{-d}$ is a direct result of the probability theory, rather than the result induced by the quantum fluctuations in QPT. For large $L$, Eq.~(\ref{eq:op4}) indicates that in the short-time stage
\begin{equation}
M^2(\tau,L)=L^{-d}\tau^{-2\beta/\nu z+d/z}(1+a_M\tau^{-\omega})^{-2\beta/\nu z+d/z}, \label{eq:op9}
\end{equation}
according to Eq.~(\ref{eq:general6}). Compared with Eq.~(\ref{eq:op2a}), again, a scaling correction factor is multiplied here. 



\begin{figure*}[tbp]
\centering
  \includegraphics[width=\linewidth,clip]{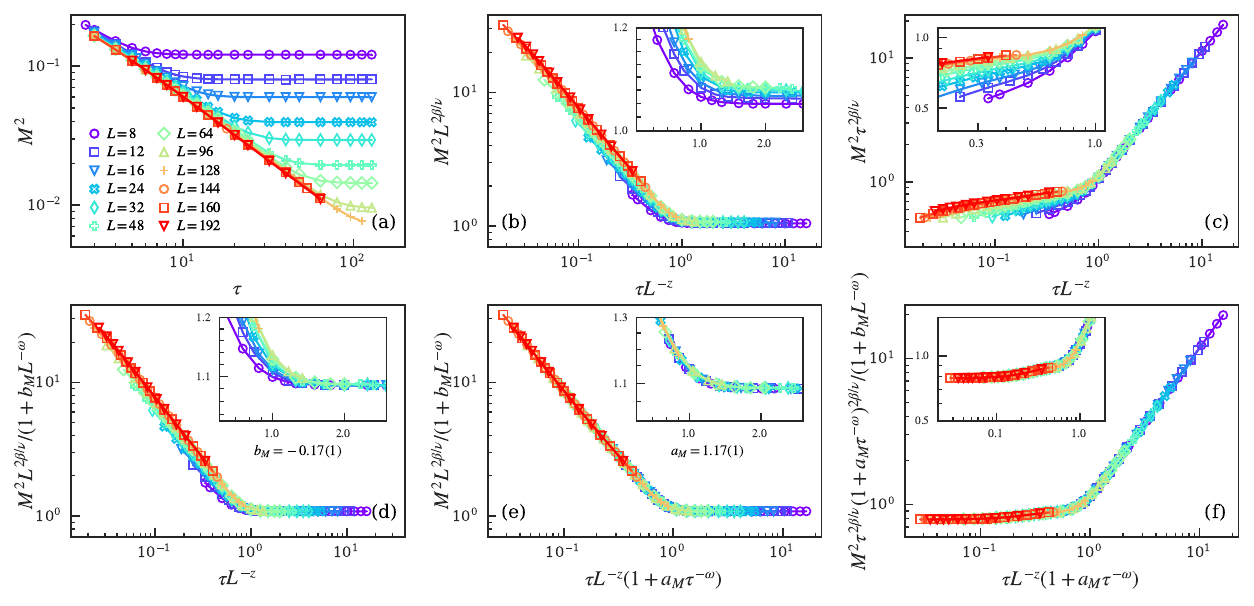}
  \vskip-3mm
  \caption{(a) Evolution of $M^2$ from the completely ordered AFM initial state for different lattice sizes at the critical point is plotted. Rescaled curves are shown in (b) and (c) according to Eqs.~(\ref{eq:op1a}) and~(\ref{eq:op1}), respectively. Rescaled curves with only finite-size scaling corrections considered are shown in (d). Rescaled curves according to Eqs.~(\ref{eq:op3}) and (\ref{eq:op4}), respectively, in which both short-time and finite-size scaling corrections are added. All insets address the details of the scaling collpase. Here both of $\beta/\nu=0.5185$ and $\omega=0.78$~\cite{Vicari2002prb,RGuida1998} are set as input. Double logarithmic scales are used.
  }
  \label{figure3}
\end{figure*}

\begin{figure}[tbp]
\centering
  \includegraphics[width=\linewidth,clip]{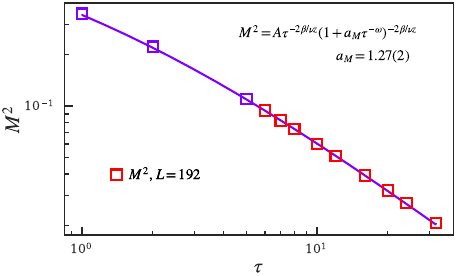}
  \vskip-3mm
  \caption{Dependence of $M^2$ on the evolution imaginary time $\tau$ with the ordered initial state for $L=192$ at the critical point. Power-law fitting according to Eq.~(\ref{eq:op5}) gives $a_M=1.27(2)$. Here both of $\beta/\nu=0.5185$ and $\omega=0.78$~\cite{Vicari2002prb,RGuida1998} are set as input. Double logarithmic scales are used.
  }
  \label{figure4}
\end{figure}

\section{\label{Results}Results}
\subsection{Numerical method}
We employ the projector QMC to implement the imaginary-time relaxation critical dynamics of model~(\ref{eq:hamiltonian}). We here briefly out line the method.

To realize the Schr\"odinger dynamics as introduced in Sec.~\ref{sectheory1}, in the projector QMC method, one takes the series expansion of the imaginary-time evolution operator $U(\tau)=\exp(-\tau H)$ in powers of $H^n$, and apply to the initial state $|\psi(0)\rangle$, giving $|\psi(\tau)\rangle=\sum_{n=0}^{\infty}\frac{\tau^n}{n!}(-H)^n|\psi(0)\rangle$. After splitting the Hamiltonian into bond operators and inserting unit operators into the operator sequence, the normalization $Z=\langle\psi(\tau)|\psi(\tau)\rangle$ can be importance sampled with the wave function written in a chosen basis, which is the standard spin-${z}$ basis or the valence-bond basis in this work, depending on the initial state desired. The actual expansion power $n$ is truncated to some maximum cut-off length that scales as $L^d\tau$ with vanishing truncation error.
A full Monte Carlo sweep of the important sampling procedure consists of local diagonal updates and global off-diagonal updates, which updates the operator sequence and the basis states simultaneously.
The local diagonal updates are carried out first, which replace unit operators with diagonal ones with appropriate acceptance rate and vice versa in the operator sequence. Then the global operator-loop updates follow up, which switch the operator types from diagonal to off-diagonal or vice versa and update the corresponding basis states. Detailed balance and ergodicity are maintained. The computational consumption of a full sweep of Monte Carlo update scales as $2\tau L^d$. The update schemes are mostly the same as those in the standard stochastic series expansion QMC method~\cite{Sandvik2010review}. Measurements are carried out in the middle of the double-sided projection, as indicated in Eq.~(\ref{eq:qtau}).
In studies of relaxation dynamics, the initial state of the system is crucial. It is important to note that in order to realize a given initial state, the imaginary-time boundaries are set fixed during the updates.
In this work, three types of initial states are considered: (i) completely ordered AFM state, (ii) random disordered state, and (iii) dimerized PM state, as shown in Fig.~\ref{figure1}. Types (i) and (ii) are implemented in the spin-$z$ basis while type (iii) employs the valence-bond basis so to maintain the dimerized order in the initial state. For a more detailed introduction of the method, we refer to the literature~\cite{Sandvik2010review,sandvik2010prb}

\subsection{\label{afm} AFM initial state}
We first investigate the imaginary-time relaxation critical dynamics from the completely ordered AFM initial state. 

The dynamics of the dimensionless Binder cumulant is shown in Fig.~\ref{figure2} (a). Figure~\ref{figure2} (b) shows that although rescaled curves of $U$ versus $\tau L^{-z}$ for different $L$ tends to collapse in comparison to Fig.~\ref{figure2} (a), apparent deviation remains for short-time and small-size cases. Then, we add the scaling corrections and rescale $U$ and $\tau$ according to Eq.~(\ref{eq:binder}). By tuning $a_U$ and $b_U$, we find the rescaled curves collapse very well, as shown in Fig.~\ref{figure2} (c). These results demonstrate the necessity of short-time and finite-size scaling corrections in the imaginary-time relaxation dynamics of model~(\ref{eq:hamiltonian}). In particular, in Fig.~\ref{figure2} (c), both the correction exponents $\omega_\tau$ and $\omega_L$ are chosen as $\omega=0.78$, which is analytically obtained in Ref.~\cite{RGuida1998} and numerically verified in Ref.~\cite{Nvsen2018prl}, confirming $\omega_\tau=\omega_L$ and the discussion below Eq.~(\ref{eq:general2}).

For the dynamic scaling behaviors of $M^2$, the evolution of $M^2$ for different system sizes is shown in Fig.~\ref{figure3} (a). By rescaling $M^2$ and $\tau$ as $M^2L^{2\beta/\nu}$ and $\tau L^{-z}$ according to Eq.~(\ref{eq:op1a}), we find in Fig.~\ref{figure3} (b) that the rescaled curves in the short-time stage have apparent discrepancies although they tend to get close to each other. In particular, in the inset of Fig.~\ref{figure3} (b), one can find the discrepancy also occurs in the equilibrium region. In addition, Fig.~\ref{figure3} (c) shows that rescaled curves of $M^2\tau^{2\beta/\nu z}$ versus $\tau L^{-z}$ according to Eq.~(\ref{eq:op1}) also do not match with each other, in particular, for short-time and small-size regions. Moreover, in the short-time stage, Fig.~\ref{figure3} (c) shows that the rescaled curves are not parallel to the horizontal axis, even for large system size, demonstrating that the scaling relation of Eq.~(\ref{eq:op1b}) does not give a complete description of the scaling behavior of $M^2$ in the short-time stage. Accordingly, scaling corrections are needed to improve the dynamic scaling theory discussed in Sec.~\ref{sectheory1}.

A natural question is whether these scaling discrepancies can be eliminated by usual finite-size scaling corrections. To examine it, in Fig.~\ref{figure3} (d), only the finite-size scaling correction is introduced. We find that for $b_M=-0.17(1)$ the rescaled curves match with each other quite well in the long-time stage, but they still deviate from each other in the short-time stage. Thus, an independent short-time scaling correction is required.

In Fig.~\ref{figure3} (e), both the short-time and finite-size scaling corrections are included according to Eq.~(\ref{eq:op3}). By tuning the coefficient before $\tau^{-\omega}$, $a_M$, but fixing the coefficient before $L^{-\omega}$, $b_M$, same as that in Fig.~\ref{figure3} (d), we find that for $a_M=1.17(1)$ the rescaled curves of $M^2L^{2\beta/\nu}$ versus $\tau L^{-z}$ can collapse quite well in the whole relaxation process.

In addition, by substituting the obtained $a_M$ and $b_M$ into Eq.~(\ref{eq:op4}) and rescaling the data according to this equation, we find that the rescaled curves also collapse quite well, as shown in Fig.~\ref{figure3} (f). These results not only successfully verify the effectiveness of scaling forms of Eqs.~(\ref{eq:op3}) and (\ref{eq:op4}), but also determine the coefficient of the short-time scaling correction. Moreover, in Fig.~\ref{figure3} (f), the rescaled curves in the short-time stage are mainly parallel to the abscissa axis, demonstrating again that appropriate scaling corrections have been established.

To further reveal the short-time dynamic scaling behavior of $M^2$, in Fig.~\ref{figure4}, we directly fit the curve of $M^2$ versus $\tau$ for large size $L=192$ according to Eq.~(\ref{eq:op5}) with the critical exponents set as input. We find that the prefactor before $\tau^{-\omega}$, $a_M$, determined from this fitting is $a_M=1.27(2)$, which is close to that obtained from data collapse in Fig.~\ref{figure3}. Accordingly, we not only confirm that in the short-time stage, the evolution of $M^2$ satisfies Eq.~(\ref{eq:op5}), but also verify the value of $a_M$.

\begin{figure*}[tbp]
\centering
  \includegraphics[width=\linewidth,clip]{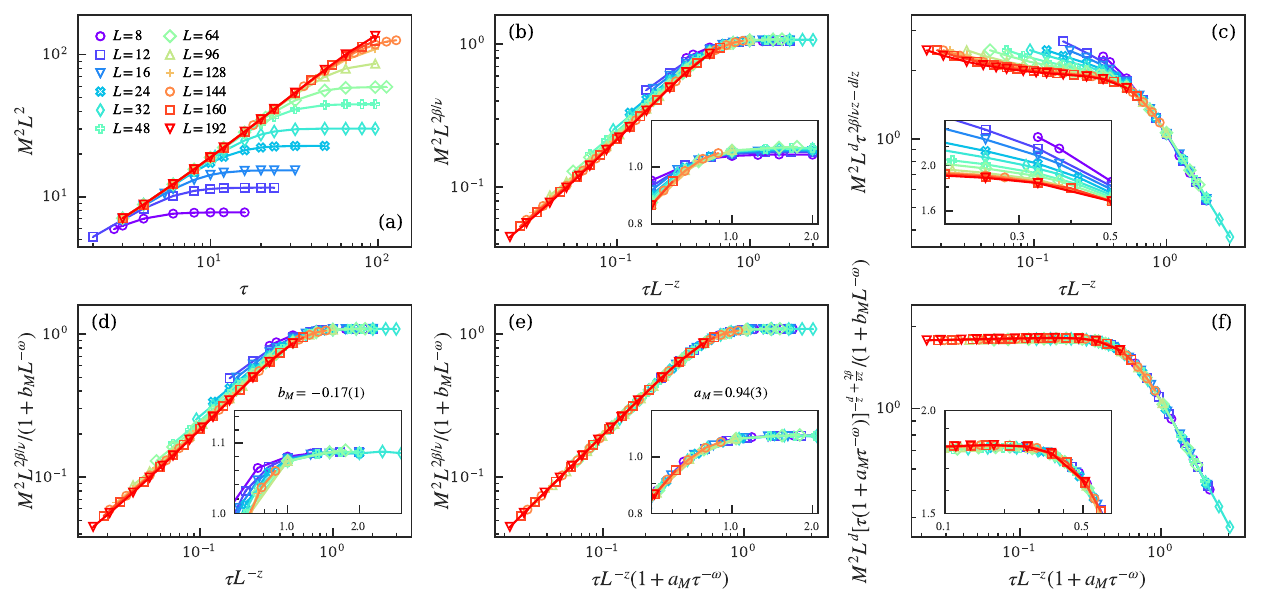}
  \vskip-3mm
  \caption{(a) Evolution of $M^2$ from the completely disordered initial state for different lattice sizes at the critical point is plotted. Rescaled curves are shown in (b) and (c) according to Eqs.~(\ref{eq:op1a}) and~(\ref{eq:op2}), respectively. Rescaled curves with only finite-size scaling corrections considered are shown in (d). Rescaled curves according to Eqs.~(\ref{eq:op3}) and (\ref{eq:op8}), respectively, in which both short-time and finite-size scaling corrections are added. All insets address the details of the scaling collpase. Here both of $\beta/\nu=0.5185$ and $\omega=0.78$~\cite{Vicari2002prb,RGuida1998} are set as input. Double logarithmic scales are used.
  }
  \label{figure5}
\end{figure*}

\begin{figure}[tbp]
\centering
  \includegraphics[width=\linewidth,clip]{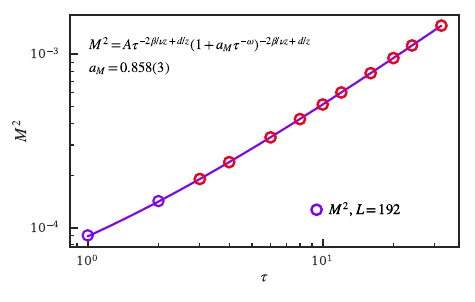}
  \vskip-3mm
  \caption{Dependence of $M^2$ on the evolution imaginary time $\tau$ with the disordered initial state for $L=192$ at the critical point. Power-law fitting according to Eq.~(\ref{eq:op9}) gives $a_M=0.858(3)$. Here both of $\beta/\nu=0.5185$ and $\omega=0.78$~\cite{Vicari2002prb,RGuida1998} are set as input. Double logarithmic scales are used.
  }
  \label{figure6}
\end{figure}

\subsection{\label{disorder} Disordered initial state}
Next, we investigate the imaginary-time relaxation critical dynamics from the completely disordered initial state. This initial state can be regarded as the high-temperature thermal state, as illustrated in Fig.~\ref{figure1}. We focus on the critical dynamics of $M^2$.

Figure~\ref{figure5} (a) shows the evolution of $M^2$ for different system sizes. Different from the decay feature for the ordered initial state, here, $M^2$ increases as $\tau$ increases. In addition, in the short-time stage, for large $L$, when the correlation length is smaller than the lattice size, $M^2\propto L^{-d}$. Thus, when plotting $M^2L^d$, we find that in the short-time and large-size region, the curves match with each other. This reflects that the relation of $M^2\propto L^{-d}$ does not need a scaling correction, since it is a direct result of the central limit theorem, as discussed in Sec.~\ref{sectheory2}.

We then rescale $M^2$ and $\tau$ for different system sizes according to the finite-size scaling form without scaling corrections, i.e., Eq.~(\ref{eq:op1a}) and show the results in Fig.~\ref{figure5} (b). From Fig.~\ref{figure5} (b) and its inset, one finds that apparent separations appear in the short-time and small-size regions. The discrepancies in the short-time stage are more obvious when $M^2$ is rescaled according to Eq.~(\ref{eq:op2}), as shown in Fig.~\ref{figure5} (c). In addition, Fig.~\ref{figure5} (c) also sdemonstrates that in the short-time stage, the evolution of $M^2$ does not satisfy Eq.~(\ref{eq:op2a}), since the rescaled curves are not parallel to the horizontal axis. Accordingly, scaling corrections are needed.

By including the finite-size scaling correction the same as the one in Fig~\ref{figure3}, Fig.~\ref{figure5} (d) shows that this scaling correction can remedy the scaling mismatching in the long-time equilibrium stage. However the discrepancy still exists in the short-time stage.

These results inspire us to include both short-time and finite-size scaling corrections, similar to the previous case with an ordered initial state. In Fig.~\ref{figure5} (e), we rescale the curves of $M^2$ versus $\tau$ for different sizes according to Eq.~(\ref{eq:op3}), then tune the coefficient $a_M$ of the short-time correction term $\tau^{-\omega}$ with $b_M$ fixed as its equilibrium value, i.e., $b_M=-0.17(1)$. We find in Fig.~\ref{figure5} (e) that for $a_M=0.94(3)$, the rescaled curves for different sizes collapse quite well in the whole relaxation process, confirming the availability of Eq.~(\ref{eq:op3}). Besides, here the value of the prefactor $a_M$ is obviously different from the one for the ordered initial state, demonstrating that this coefficient depends on the initial state.

In addition, with the obtained $a_M$ and $b_M$, we rescale the data according to Eq.~(\ref{eq:op8}), we find that the rescaled curves also collapse quite well, as shown in Fig.~\ref{figure5} (f). These results successfully verify the effectiveness of the short-time scaling corrections in Eqs.~(\ref{eq:op3}) and (\ref{eq:op8}). Moreover, as shown in Fig.~\ref{figure5} (f), the rescaled curves almost keep aclinic in the short-time stage, demonstrating again that appropriate scaling corrections have been introduced.

\begin{figure*}[tbp]
\centering
  \includegraphics[width=\linewidth,clip]{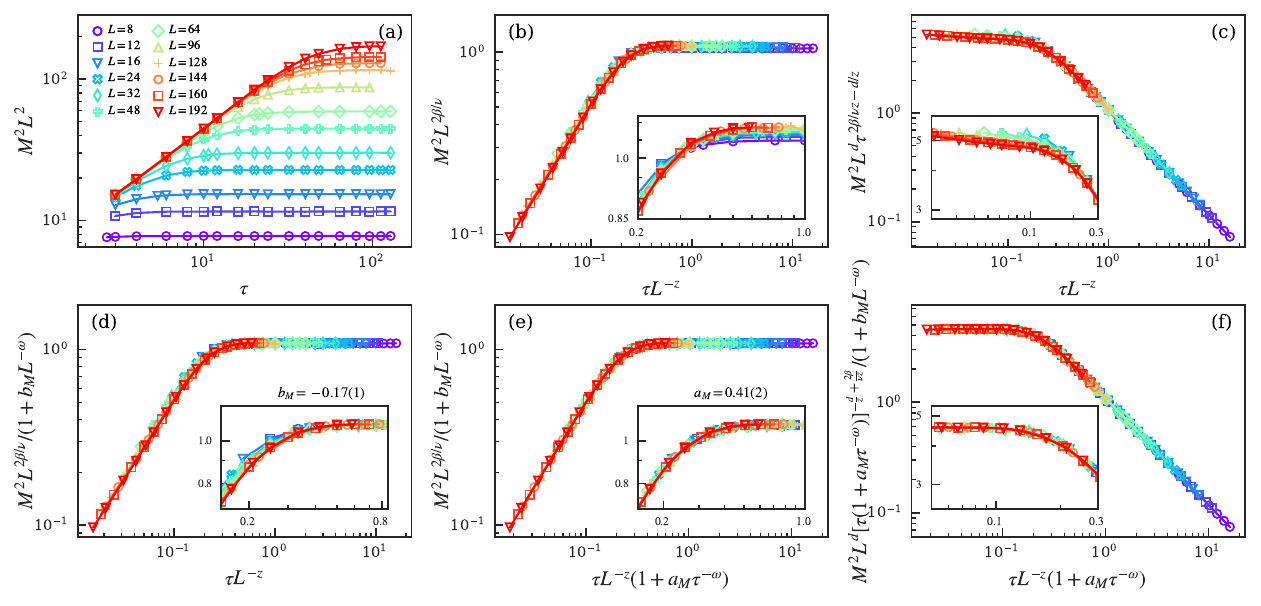}
  \vskip-3mm
  \caption{(a) Evolution of $M^2$ from the quantum PM initial state for different lattice sizes at the critical point is plotted. Rescaled curves are shown in (b) and (c) according to Eqs.~(\ref{eq:op1a}) and~(\ref{eq:op2}), respectively. Rescaled curves with only finite-size scaling corrections considered are shown in (d). Rescaled curves according to Eqs.~(\ref{eq:op3}) and (\ref{eq:op8}), respectively, in which both short-time and finite-size scaling corrections are added. All insets address the details of the scaling collpase. Here both of $\beta/\nu=0.5185$ and $\omega=0.78$~\cite{Vicari2002prb,RGuida1998} are set as input. Double logarithmic scales are used.
  }
  \label{figure7}
\end{figure*}

\begin{figure}[tbp]
\centering
  \includegraphics[width=\linewidth,clip]{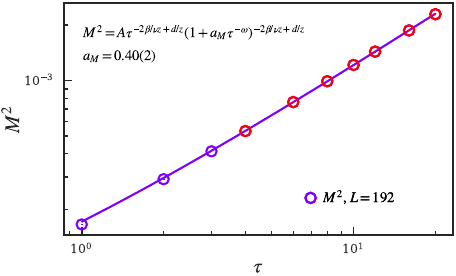}
  \vskip-3mm
  \caption{Dependence of $M^2$ on the evolution imaginary time $\tau$ with the PM initial state for $L=192$ at the critical point. Power-law fitting according to Eq.~(\ref{eq:op9}) gives $a_M=0.40(2)$. Here both of $\beta/\nu=0.5185$ and $\omega=0.78$~\cite{Vicari2002prb,RGuida1998} are set as input. Double logarithmic scales are used.
  }
  \label{figure8}
\end{figure}

To further reveal the short-time dynamic scaling behavior of $M^2$, we directly fit the curve of $M^2$ versus $\tau$ for large size $L=192$ according to Eq.~(\ref{eq:op9}) with the critical exponents set as input. We find that the prefactor of the short-time correction obtained from this fitting is $a_M=0.858(3)$, close to that obtained from data collapse in Fig.~\ref{figure5}. Accordingly, we not only show that in the short-time stage, the evolution of $M^2$ satisfies Eq.~(\ref{eq:op9}), but also confirm the value of the value of the coefficient of the short-time correction.

\subsection{\label{para} Paramagnetic initial state}
In this section, we turn to investigate the evolution of $M^2$ from the quantum PM initial state, as shown in Fig.~\ref{figure1}. Although apparently different from the thermally disordered state, this PM state is also magnetically disordered with zero magnetization. Therefore, one expects similar scaling behaviors of $M^2$ to those in Sec.~\ref{disorder}.

Figure~\ref{figure7} (a) shows the evolution of $M^2$ for different system sizes. Apparent discrepancies can be found in the rescaled curves when they are rescaled according to the scaling functions Eqs.~(\ref{eq:op1a}) and (\ref{eq:op2}) without scaling corrections, as shown in Figs.~\ref{figure7} (b) and (c). With only finite-size scaling correction included, Fig.~\ref{figure7} (d) shows that the discrepancy still exists in the short-time region.

Then, as shown in Figs.~\ref{figure7} (e) and (f), by rescaling the curves of $M^2$ versus $\tau$ for different sizes according to Eqs.~(\ref{eq:op3}) and (\ref{eq:op8}), respectively, we find the rescaled curves for different sizes collapse well in the whole relaxation process when the prefactor $a_M$ in the short-time correction is chosen as $a_M=0.41(2)$. These results confirm the universality of the scaling forms of Eqs.~(\ref{eq:op3}) and (\ref{eq:op8}). Moreover, as shown in Fig.~\ref{figure7} (f), the rescaled curves almost keep parallel to the horizontal axis in the short-time stage, demonstrating again that appropriate scaling corrections have been built.

The short-time dynamic scaling behavior of $M^2$ with the PM initial state is further explored in Fig.~\ref{figure8}. Therein we directly fit the curve of $M^2$ versus $\tau$ for large size $L=192$ according to Eq.~(\ref{eq:op9}) with the critical exponents set as input. We find that the short-time correction prefactor is $a_M=0.40(2)$, close to that obtained from data collapse in Fig.~\ref{figure7}. Accordingly, we not only show that in the short-time stage, the evolution of $M^2$ satisfies Eq.~(\ref{eq:op8}), but also confirm the value of $a_M$.

\section{\label{summary} Summary}
In summary, we have studied the imaginary-time relaxation dynamics in the $2$D dimerized Heisenberg model. We have shown that remarkable discrepancies are found when the imaginary-time critical relaxation behaviors are described by the usual scaling forms. Moreover, we have found that besides the finite-size scaling correction, an additional short-time scaling correction is required to be included in the dynamic scaling theory. A full scaling form, including both short-time and finite-size scaling corrections, has been proposed. From this scaling form, modified short-imaginary-time relaxation scaling properties have been obtained. We have then verified these full scaling forms and short-time scaling properties for different initial states via QMC simulations. Note that the imaginary-time dynamics have been realized experimentally in the platforms of quantum computers to prepare the ground state of quantum systems~\cite{Motta2020naturephyscis,Nishi2021njp,Pollmann2021prxq}. In particular, the short-imaginary-time scaling behavior is also been found in these systems~\cite{Zhang2023}. Thus, it is expected that our results could be verified in the near-term quantum devices. Moreover, although the real-time dynamics have unitary nature, which is quite different from the dissipative nature of imaginary-time dynamics, both of real and imaginary time share the same scaling dimension~\cite{Yin2019prl,Mitra2015prb,Halimeh2023prb,Marino2022review,JianLi2023arxiv}. Therefore, it is expected that the real-time critical dynamics can have similar correction forms in time direction and this work is still in progress.

\section*{Acknowledgments}
{\it Acknowledgments}--- J. Q. Cai, X. Q. Rao, and S. Yin are supported by the National Natural Science Foundation of China (Grants No. 12075324 and No. 12222515). S. Yin is also supported by the Science and Technology Projects in Guangdong Province (Grants No. 211193863020). Y.-R. Shu~is supported by the National Natural Science Foundation of China, Grant No.~12104109 and Key Discipline of Materials Science and Engineering, Bureau of Education of Guangzhou, Grant No.~202255464.
\bibliographystyle{apsrev4-1}
\bibliography{stheisenberg}
\end{document}